\documentclass[prb,aps,twocolumn,showpacs,bibnotes,superscriptaddress,epsf]{revtex4-1}

\usepackage[T1]{fontenc} 
\usepackage{graphicx}

\usepackage[urlcolor=blue]{hyperref}
\hypersetup{backref, colorlinks=true, linkcolor=blue, citecolor=blue}

\begin{document}

\title{Order-disorder transition in $p$-oligophenyls}

\author{Kai Zhang}
\affiliation{Key Laboratory of Materials Physics, Institute of Solid State Physics, Chinese Academy of Sciences, Hefei 230031, China}
\affiliation{University of Science and Technology of China, Hefei 230026, China}
\affiliation{Center for High Pressure Science and Technology Advanced Research, Shanghai 201203, China}

\author{Ren-Shu Wang}
\affiliation{Center for High Pressure Science and Technology Advanced Research, Shanghai 201203, China}

\author{Xiao-Jia Chen}\email{xjchen@hpstar.ac.cn}
\affiliation{Center for High Pressure Science and Technology Advanced Research, Shanghai 201203, China}

\begin{abstract}
   Poly($para$-phenylene) has been recognized as one important family of conducting polymers upon doping with donors or acceptors. This system possesses a chain-like structure with infinite benzene rings linked with the single C-C bond. Oligophenyls as models of poly($para$-phenylene) with short chains in the \emph{para} position were found to exhibit superconductivity at transition temperatures ranging from 3 K to 123 K upon dopant. Structural studies have revealed that there exist the order-disorder transitions in many \emph{p}-oligophenyls with almost doubled lattice constants in the \emph{b} and \emph{c} directions at low temperatures, seemingly supporting the formation of the charge-density-wave order. Such a transition is of relevance to the understanding of the emergence of novel quantum functionality where unconventional polaronic interactions are relevant for the new emerging theory of superconductivity in this system. However, the accurate temperatures for the order-disorder transitions amongst these \emph{p}-oligophenyls are still needed to be determined. The chain length effects on the order-disorder transitions in this system remain unknown. Here we report the systematic investigation of the evolution of vibrational properties of the crystalline \emph{p}-oligophenyls over a wide temperature range. The order-disorder transition is identified by three indicators, the lowest energy peak together with the intensity ratios between the 1280 cm$^{-1}$ and 1220 cm$^{-1}$ modes and between the two modes at around 1600 cm$^{-1}$. A phase diagram of the order-disorder transition temperature as well as the melting curve for \emph{p}-oligophenyls is thus established. The former is found to increase with the chain length and saturates at around 350 K for poly($para$-phenylene).
\end{abstract}


\maketitle

\section{Introduction}

Benzene and its derivatives are always attractive materials because of the novel and excellent physical and chemical properties benefited from the delocalized cyclic continuous $\pi$ bond between the carbon atoms. Among the benzene derivatives, poly($para$-phenylene) (PPP) materials draw extensive attentions. These materials all have the structure with a certain number of phenyl rings connecting at the \emph{para} positions. The short \emph{p}-oligophenyls (POPs) are named with the prefixes of bi-, ter-, quater-, quinque-, and sexi-, before the phenyl for the two to six phenyl rings compounds, respectively. Early studies revealed that these materials all show insulating or semiconducting behavior in the normal state, whereas their conductivities increase by several orders of magnitude due to the formations of the polarons and bipolarons after alkali or alkali earth metals doping.\cite{Havin,Ivory,Shack} The molecular structure in the doped compounds changes from benzene structure to quinoid structure, it is available to the conduction electrons so that the localized electrons in one molecule, especially the $\pi$-electrons, can transfer to the next molecules with combining to the doped atoms.\cite{Zanno,Cuff-0} In addition, the conductivities of the doped PPP materials show an increasing tendency with increasing molecular chain length.\cite{Havin,Shack} Such versatile properties of PPP materials make them to be extensively applied in the scientific research and applications. For example, the biphenyl (P2P) can be a starting material of a host of organic compounds, the \emph{p}-terphenyl (P3P) is often used as an ultraviolet laser dye,\cite{Cecco} the \emph{p}-quaterphenyl (P4P) and its derivatives are always extensively applied in organic thin film transistors and organic light emitting diodes,\cite{Gundl,Hosok,Quoch,Lu} and so on. Now, the PPP materials are back to spotlight because of the recent discoveries of the superconducting phases with the transition temperatures of 3.4 K, 7.2 K, and even 123 K,\cite{Zhong-0,Huang-0,Yan-0,Wang-1,Wang-2,Wang-3,Wang-4,Mazzi} the latter is higher that the boiling point of liquid nitrogen. These are significant findings after the prediction of the superconductivity in solid benzene molecular crystal.\cite{Zhong-1} And the 7.2 K phase seemingly considers as a common phase in PPP materials. These findings make PPP materials good candidates for finding the superconductor with high transition temperatures.

The herringbone type structure also generates two adverse forces,\cite{Carre,Soto} one is the steric repulsive force resulted from the four hydrogen atoms of the neighboring two phenyl rings, and the other is the attractions among the delocalized $\pi$-electrons which tends to parallel each ring in one molecule. Due to the competing two forces, the neighboring two rings in one individual molecule, such as in the gas state or in solutions, will exhibit a stable angle. Such as the liquid biphenyl, the tilt angle in the individual molecule is confirmed to be 44 degree.\cite{Almen} However, after crystalizing, the twist angles turn into zero on average despite the strong thermal librations of the phenyl rings around the long molecular axes.\cite{Charb,Baudo-1,Baudo-2,Delug} Interestingly, it seems as if the shaky molecules of PPP materials freeze up at certain temperatures, and the interring tilt angles between rings simultaneously reemerge.\cite{Charb,Baudo-1,Baudo-2,Delug,Barba} Furthermore, the angles of the adjacent molecules have the opposite orientations. As a result, the structure of PPP materials transforms from monoclinic to triclinic with remaining four molecules in one-unit cell. Such a structural transition can be confirmed by broad humps in the specific heat measurements.\cite{Atake,Chang,Saito,Saito-1,Atake-1,Saito-2} These transitions are divided into two categories: "displacive" and "order-disorder" types. Biphenyl is the unique "displacive"-type material in PPPs.\cite{Caill,Baudo-1} It has two incommensurate structural transitions occurring at 40 K and 21 K with the reciprocal lattice vectors q$_{\delta}$=$\delta$$_a$a$^*$+$\frac{1}{2}$(1-$\delta$$_b$)b$^*$ and q$_{\delta}$=$\frac{1}{2}$(1-$\delta$$_b$)b$^*$, respectively.\cite{Bree,Atake,Caill-3,Caill-1,Atake-1,Saito-2} The other PPP materials all belong to the "order-disorder" type with the transition temperatures range from 193 K to 295 K for \emph{p}-terphenyl to \emph{p}-sexiphenyl.\cite{Baker} Raman spectroscopy as a convenient and effective tool is broadly applied to study the vibrational properties of PPP materials. The chain length effects of POPs have been investigated by Raman scattering.\cite{Zhang-1} And the Raman spectra of biphenyl, \emph{p}-terphenyl, and \emph{p}-quaterphenyl at low temperature are also performed to understand the temperature effects on the vibrational properties.\cite{Bree,Fried,Bruce,Girar,Bolto,Girar-1,Zhang,Zhang-2,Zhang-3} The peaks of those materials all exhibit drastic splits at low temperature due to the decreases of the full weight at half maximum (FWHM) below the transition temperatures. The FWHMs of the lowest energy peak and the intensity ratios of the modes at around 1220, 1280, and 1600 cm$^{-1}$ all show anomalous tendencies at around the transition temperatures. These indicate that this structural transition significantly impacts the vibrational properties of PPP materials. However, the low-temperature Raman spectra of \emph{p}-quinquephenyl (P5P) and \emph{p}-sexiphenyl (P6P) are rarely performed. Whether those remarkable indicators are applicable to the PPP materials with long chain length, it is remaining shrouded in mystery. Since the common superconducting phase was found in PPP materials, the fundamental properties of them are becoming increasingly essential and important.

Here, we present the Raman spectra of benzene, $p$-quinquephenyl, and $p$-sexiphenyl with the temperature range from 5 K to the temperature above the ambient condition. The temperature effects on the vibrational properties of the modes ranging from the lattice vibrations to the high-energy C-H vibrations, especially the lowest energy peak and the modes at around 1220, 1280, and 1600 cm$^{-1}$, are studied in detail. The results are then compared with previous works focused on the biphenyl, $p$-terphenyl, $p$-quaterphenyl.\cite{Zhang-2,Zhang,Zhang-3} Our work provides the basis to the understanding of the vibrational properties of PPP materials, and it is also of great significance for the further study on PPP materials.

\section{EXPERIMENTAL DETAILS}

In the experiments all the high-purity samples were purchased from Sinopharm Chemical Reagent and Alfa Aesarare. They were all sealed in quartz tubes with high transmittances for Raman-scattering experiments in a glove box with the moisture and oxygen levels less than 0.1 ppm. The tubes were stuck on the heater by using the glue with high thermal conductivity to reduce the error of the internal temperature. Then the heater was put in a cryogenic vacuum chamber for cooling down to low temperature. Due to the different melting points, the benzene was measured from 5 K to 300 K, the \emph{p}-quinquephenyl was measured from 5 K to 350 K, the \emph{p}-sexiphenyl was measured from 5 K to 360 K. The applied wavelength of the exciting laser was 660 nm. The power was less than 1 mW before a $\times$20 objective to avoid possible damage of samples. The integration time is 20 s. A 1024-pixel Charge Coupled Device designed by Princeton was used to recorded the Raman spectra.

\section{RESULTS AND DISCUSSION}

\begin{figure}[htbp]
\includegraphics[width=0.48\textwidth]{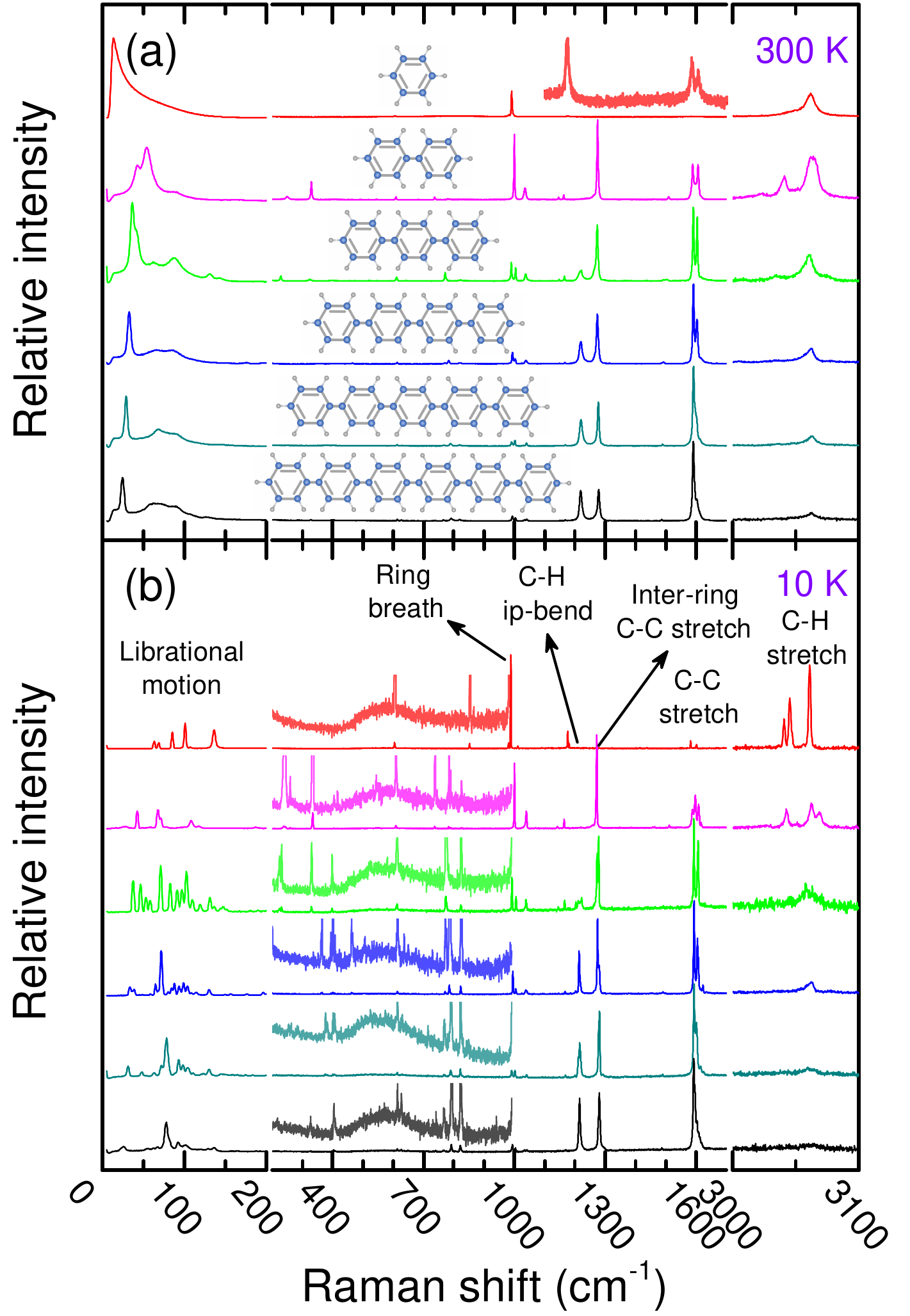}
\caption{Raman spectra of benzene, biphenyl, \emph{p}-terphenyl, \emph{p}-quaterphenyl, \emph{p}-quinquephenyl and \emph{p}-sexiphenyl measured at 300 K (a) and 10 K (b) excited by 660 nm laser. Each spectrum is plotted in the same scale. The molecular graphs of each POP material are painted in (a). Classifications of the vibration modes are depicted above the spectrum of benzene at 10 K. The partial spectra are zoomed in for clear distinction.}
\end{figure}

\begin{figure*}[htbp]
\includegraphics[width=0.95\textwidth]{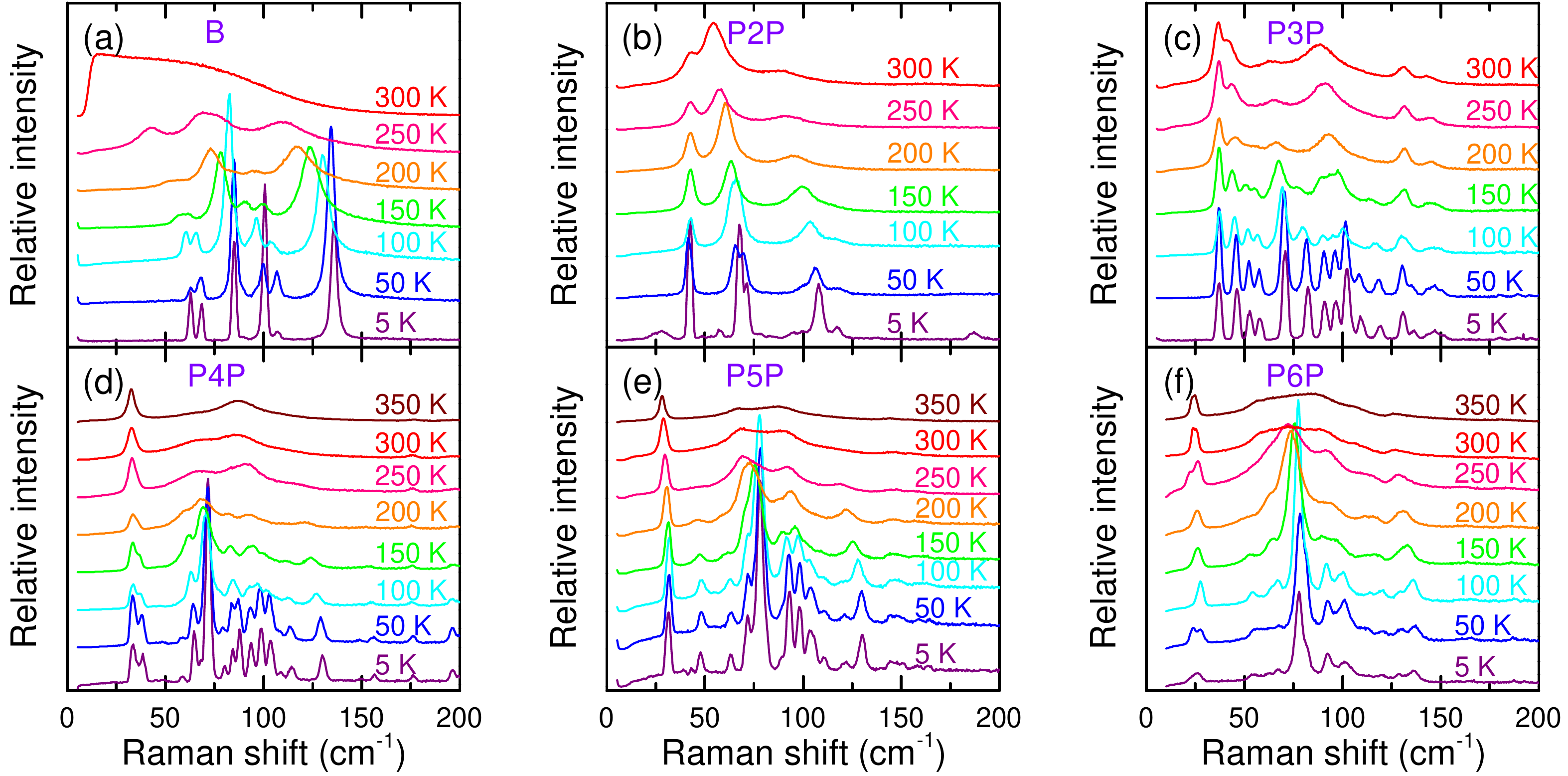}
\caption{Raman spectra of benzene, biphenyl, \emph{p}-terphenyl, \emph{p}-quaterphenyl, \emph{p}-quinquephenyl and \emph{p}-sexiphenyl measured at different temperatures. All the spectra are displaced vertically for clarification.}
\end{figure*}

Figure 1 presents the Raman spectra of the benzene and POPs from biphenyl to \emph{p}-sexiphenyl at 300 K and 10 K excited by a 660 nm laser. Some areas are zoomed in for clarifying, and the classifications of the vibrational modes are presented in Fig. 1(b).\cite{Honda} All the scattering intensities have been normalized by I$\left( \omega  \right)$ = ${I_0}\left( \omega  \right)/\left[ {n\left( {\omega ,T} \right) + 1} \right]$. Here, $n\left( {\omega ,T} \right)$ is the Bose-Einstein distribution function evaluated at mode energy $\omega$ and temperature $T$, and $I$$_0$($\omega$) is the observed intensity.\cite{Zhang-3} The internal normal modes can be approximately separated into the inter- and intramolecular terms. The former stems from the relative rotation and translation motions of the phenyl groups, and are always located below about 200 cm$^{-1}$. The latter is associated with the in-plane and out-plane vibrations in the individual molecule, and always has high energies. Due to the liquid state of benzene at ambient condition, the spectrum of benzene is very weak at 300 K and the librational modes are submerged in the strong background signal. The liquid benzene crystallizes to an orthorhombic phase at around 280 K, and the spectrum intensity becomes stronger. The five POPs belong to the monoclinic system at ambient conditions, and they all exhibit structural transitions with the structural transformation to triclinic at low temperatures. The transition temperatures are 40, 193, 233, 264, and 295 K confirmed by heat capacity measurements respectively for the P2P, P3P, P4P, P5P, and P6P.\cite{Saito-1} As seen in Fig. 1, the spectra of each material at 10 K show significant differences as compared with those at 300 K. The major features are summarized as follows:

\begin{figure*}[htbp]
\includegraphics[width=0.95\textwidth]{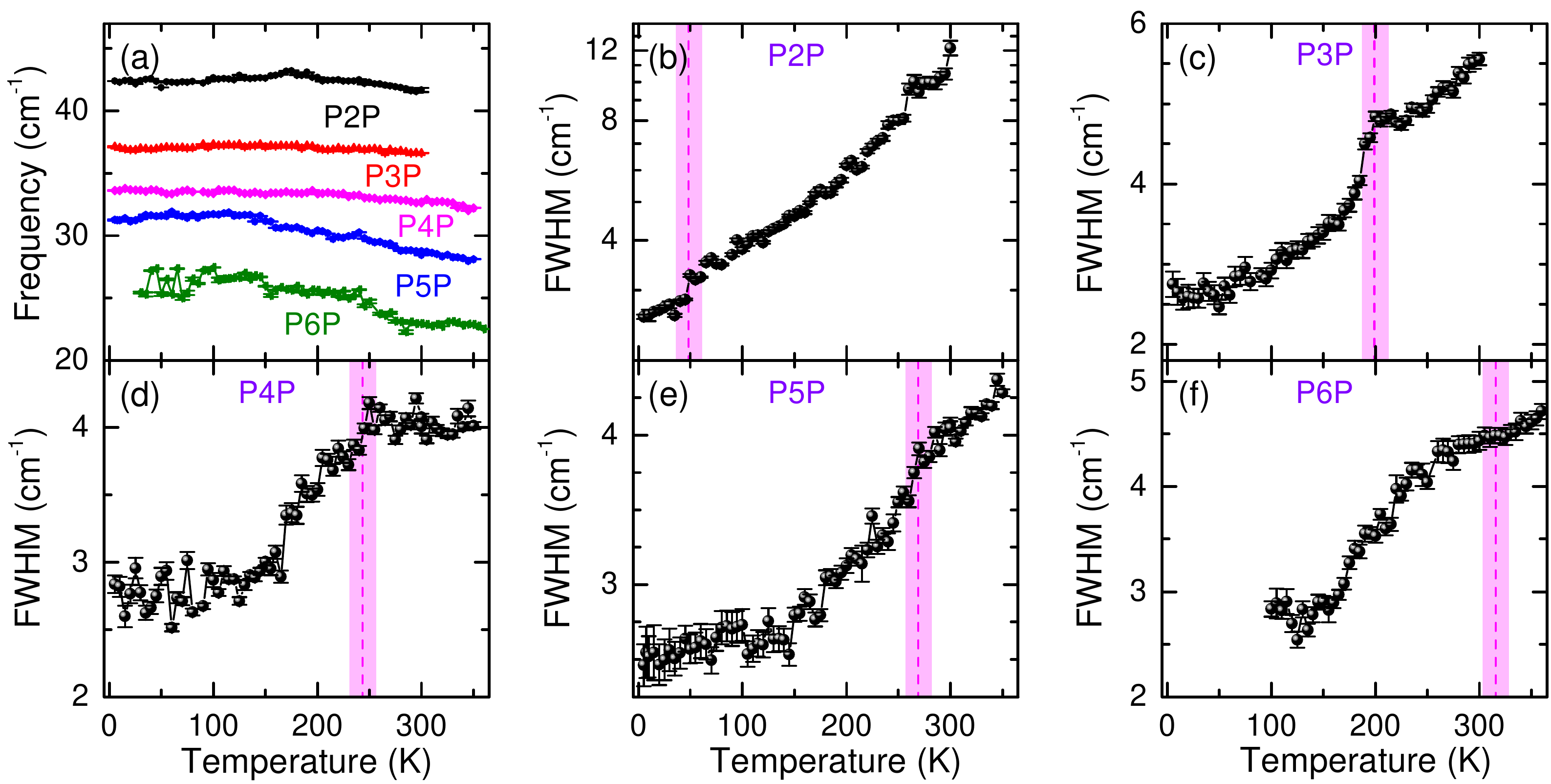}
\caption{(a) Frequencies of the lowest energy peaks of the POPs as a function of temperature. (b), (c), (d), (e), and (f) FWHMs of biphenyl, \emph{p}-terphenyl, \emph{p}-quaterphenyl, \emph{p}-quinquephenyl, and \emph{p}-sexiphenyl as a function of temperature. The structural transition temperatures are pointed out by the vertical dashed lines, and the shaded parts indicate the error ranges.}
\end{figure*}

\begin{itemize}
  \item No distinguishable peaks can be observed in the lattice motion range of the liquid benzene, whereas some peaks appear in the high energy range. This indicates that the latter peaks are associated with the vibrations among the inner atoms of the benzene ring. In addition, there are some new peaks appeared in the spectra of POPs, such as the peaks located at around 800 cm$^{-1}$,1220 cm$^{-1}$, 1280 cm$^{-1}$, 1500 cm$^{-1}$, and so on. These peaks should result from the interaction of the phenyl rings.
  \item Near all the modes show drastic splits at 10 K, especially the librational motions. The anomalous splits were confirmed by the previous Raman scattering measurements and explained resulting from the drastic decreases of the peak widths below the structure transition temperatures.\cite{Zhang-2,Zhang,Zhang-3}
  \item The tendencies of the intensities of the two ring breathing modes almost keep the same at low temperature which decrease with increasing the phenyl numbers.
  \item The modes around 1220 and 1280 cm$^{-1}$ are associated with the C-H in-plane bending modes and the interring C-C stretching modes, they stem from the interaction of the neighboring rings, thus they disappear in the spectra of benzene. The 1220 cm$^{-1}$ mode gradually decreases the intensity with decreasing chain length and it absolutely vanishes in P2P, whereas the 1280 cm$^{-1}$ have less changes. This scenario agrees with the previous work which concludes that the former mode is more sensitive to the chain length and planarity than the latter. The modes ar around 1600 cm$^{-1}$ are associated with the C-C stretching modes, and the intensities of these modes increase with increasing phenyl numbers. They all exhibit severe splits at low temperatures.
  \item The backgrounds of all the materials are nearly straight at ambient condition, whereas they raise as a broad hill. In addition, peaks located at around 500 cm$^{-1}$ with the width more than 100 cm$^{-1}$ appear at low temperatures in the six materials. They have been excluded from the external dirty signals. These anomalous phenomena should result from the increase of the fluorescence effect.\cite{Zhang-2,Zhang,Zhang-3}
  \item The intensities of the high-energy C-H stretching modes decrease as the phenyl rings number is increased, and they also dramatically split at low temperatures. This behavior implies that these peaks are associated from the terminal rings, or it can be occurred due to the monotonous decrease of the H and C ratios with increasing the chain length.\cite{Zhang-1}
\end{itemize}

The detailed Raman spectra with respect to the lattice motions of each material are presented in Fig. 2, the spectra of the P2P, P3P, and P4P are cited from the published references.\cite{Zhang-2,Zhang,Zhang-3} The most significant feature is the drastic splits of the librational modes with decreasing temperature. The splits have been confirmed to result from the decrease of the FWHMs at low temperatures.\cite{Zhang-2,Zhang,Zhang-3} Otherwise, all the peaks have blue-shifts, and the intensities simultaneously increases. For benzene, the lattice librational peaks suddenly appear after crystallizing, and then blueshift upon cooling. The POPs show entirely different Raman spectra from benzene due to the different crystal structures. Meanwhile, among the five POPs, the spectra of the P2P also show a little distinct. First, due to the incommensurate lattice modulation in P2P, several new peaks appear below the transition temperature, whereas no peaks can be observed in the spectra of other POPs without any modulations on the lattice. Second, the wavenumbers of the peaks in P2P, as well as the peak numbers at low temperatures, are different with comparing to other POPs even though they belong to the same crystal structure (monoclinic in the normal state and triclinic after the structural transition). Such phenomena should result from the strong intermolecular forces which are comparable to some intramolecular forces.\cite{Burgo} That is also the reason why the unique displacive€ transition of P2P comes from. In addition, on can also see that the intensities of those peaks at 5 K decrease with increasing phenyl ring numbers from 3 to 6 despite the transition temperatures gradually increase. This should result from the gradual increase of the lattice parameters, the interaction forces among molecules decrease as the chain length is increased.

\begin{figure*}[htbp]
\includegraphics[width=0.95\textwidth]{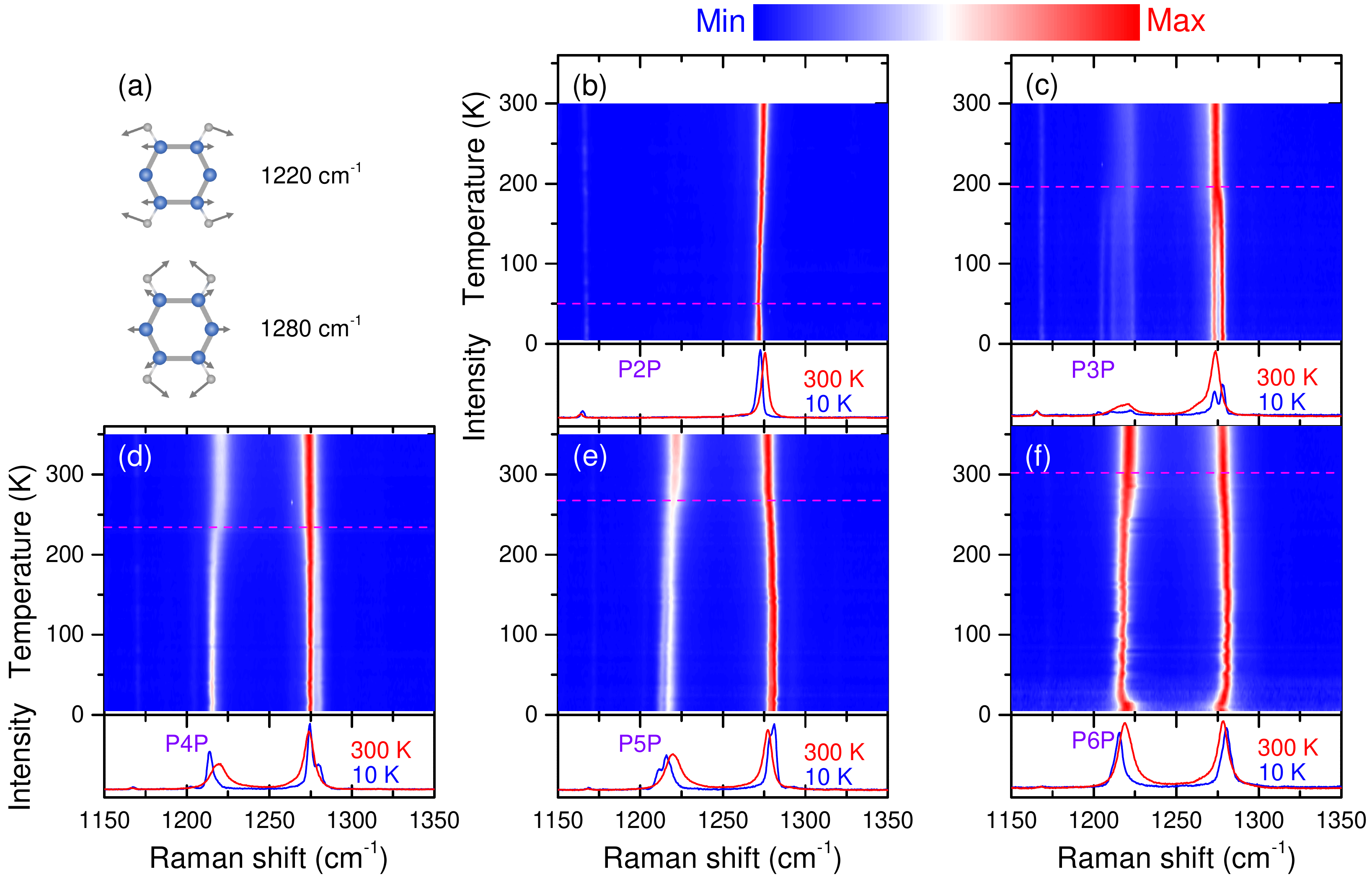}
\caption{(a) The displacement patterns of the 1220 and 1280 cm$^{-1}$ modes. (b)-(f) Temperature dependent intensity maps of the biphenyl, \emph{p}-terphenyl, \emph{p}-quaterphenyl, \emph{p}-quinquephenyl and \emph{p}-sexiphenyl. Raman spectra at 5 K and 300 K of each POPs are depicted for clarification. The transition temperatures are pointed out by the horizontal dashed lines.}
\end{figure*}

\begin{figure*}[htbp]
\includegraphics[width=0.95\textwidth]{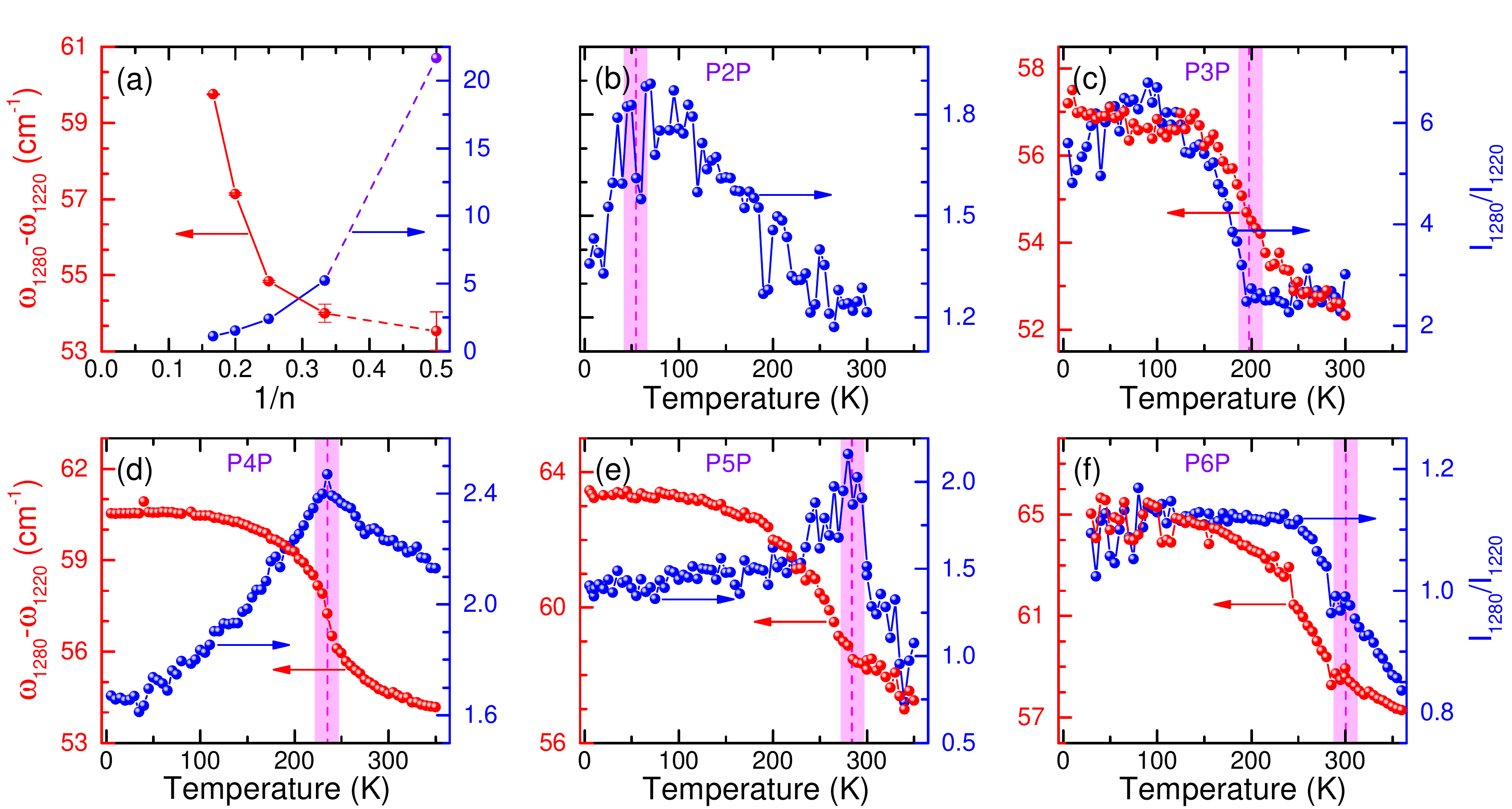}
\caption{(a) Intensity ratios and the energy separations of the 1220 and 1280 cm$^{-1}$ modes as a function of phenyl number. The intensity ratio of the biphenyl is an estimated value due to the weak intensity of the 1220 cm$^{-1}$ mode, and thus plots by a violet solid ball. (b)-(f) Intensity ratios and energy separations of the 1220 and 1280 cm$^{-1}$ modes as a function of temperature of POPs materials.}
\end{figure*}

\begin{figure*}[htbp]
\includegraphics[width=0.95\textwidth]{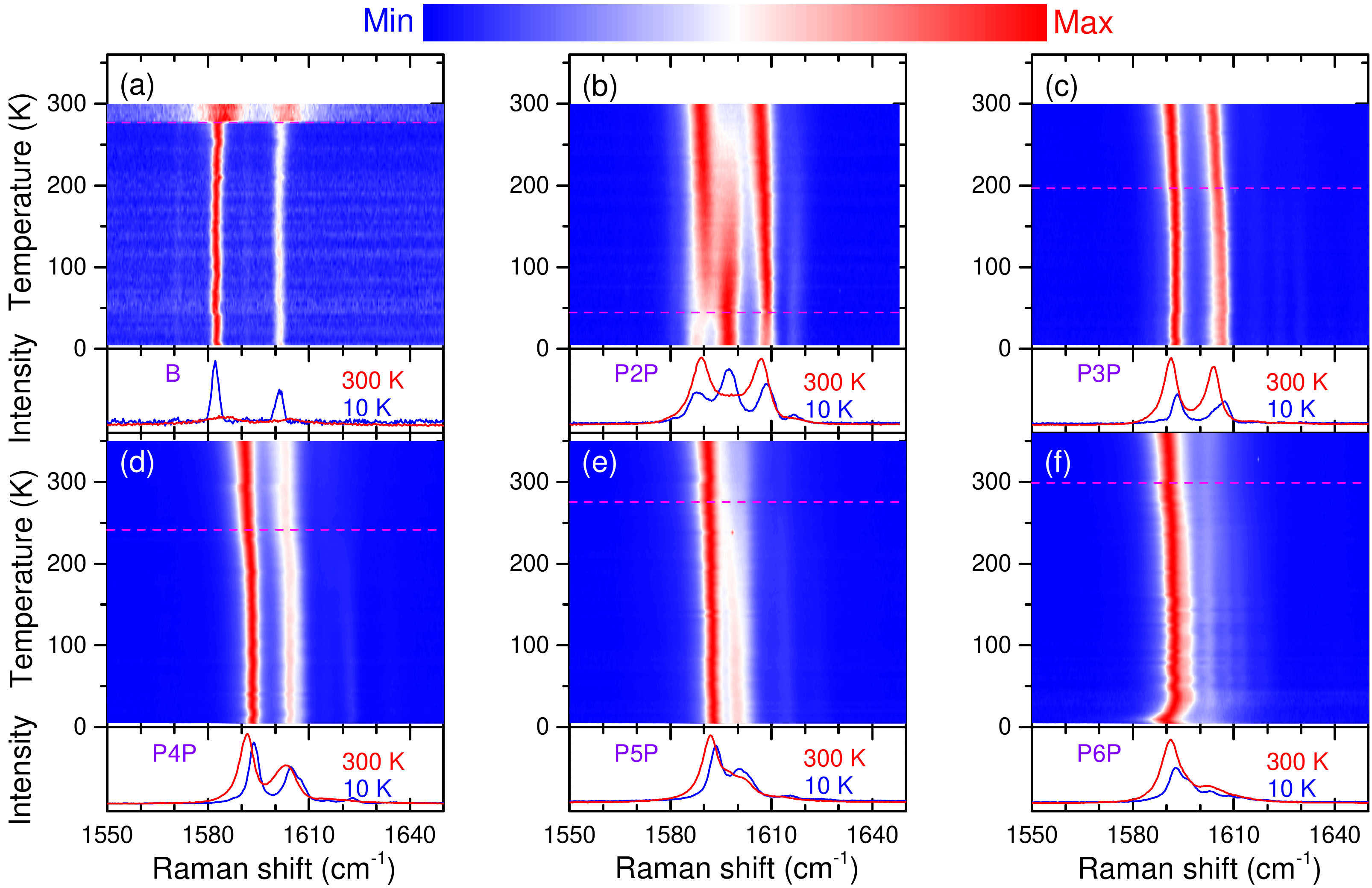}
\caption{Temperature dependent intensity maps of the benzene, biphenyl, \emph{p}-terphenyl, \emph{p}-quaterphenyl, \emph{p}-quinquephenyl and \emph{p}-sexiphenyl. Raman spectra at 5 K and 300 K of each POPs are depicted for clarification. The transition temperatures are pointed out by the horizontal dashed lines.}
\end{figure*}

Among the lattice modes, the lowest frequency mode draws a lot of attentions. It can be a good indicator of the chain length at normal state due to the regular tendency of the frequency as the dependence of the molecular chain length. Here, we fit the lowest frequency peaks of each material by a Lorentz function, and the temperature dependent frequencies and FWHMs are shown in Fig. 3. Indeed, the frequency of this peak monotonously decreases with increasing chain length. This tendency also stands at low temperatures even though all the lowest energy peaks of the POPs blueshift with decreasing temperature (see Fig. 3(a)). Thus, the lowest energy peak is adequate to be a good indicator of the molecule length in PPP materials at different temperatures. The temperature dependent FWHMs of the POPs are presented in Figs. 3(b)-3(f). Previous Raman scattering works discovered that the drastic peak splits at low temperatures mainly result from the abrupt reductions of the peak widths at around the structural transition temperature, and the lowest energy peak is a typical example.\cite{Zhang-1,Zhang-2,Zhang,Zhang-3} As seen in Fig. 3, all the POPs accord with this scenario, including the "displacive" type biphenyl which also exhibits a width drop at around the transition temperature. More interestingly, for the "order-disorder"-type POPs, the magnitudes of the FWHM gradually decrease with increasing chain length, the magnitude almost vanishes for \emph{p}-sexiphenyl. Such a tendency should result from the increasing chain length. With increasing chain length, the thermal dynamics among the phenyl rings become stable. Thus, the twist angles between the neighboring phenyl rings reduce with increasing molecular length, and the infinite long polymer is believed planar at various conditions. Consequently, the influence of the reconstructed angle below the transition temperature gradually fades away. In addition, the transition temperatures confirmed by Raman scattering are all higher than those decided by heat capacity measurements. Such a condition results from the entrance of the thermal anomalous state which has been discovered by thermodynamic studies.\cite{Atake,Chang,Saito,Saito-1,Atake-1,Saito-2} This also indicates that the internal vibrational properties are very sensitive to the structure changes.

\begin{figure*}[htbp]
\includegraphics[width=0.95\textwidth]{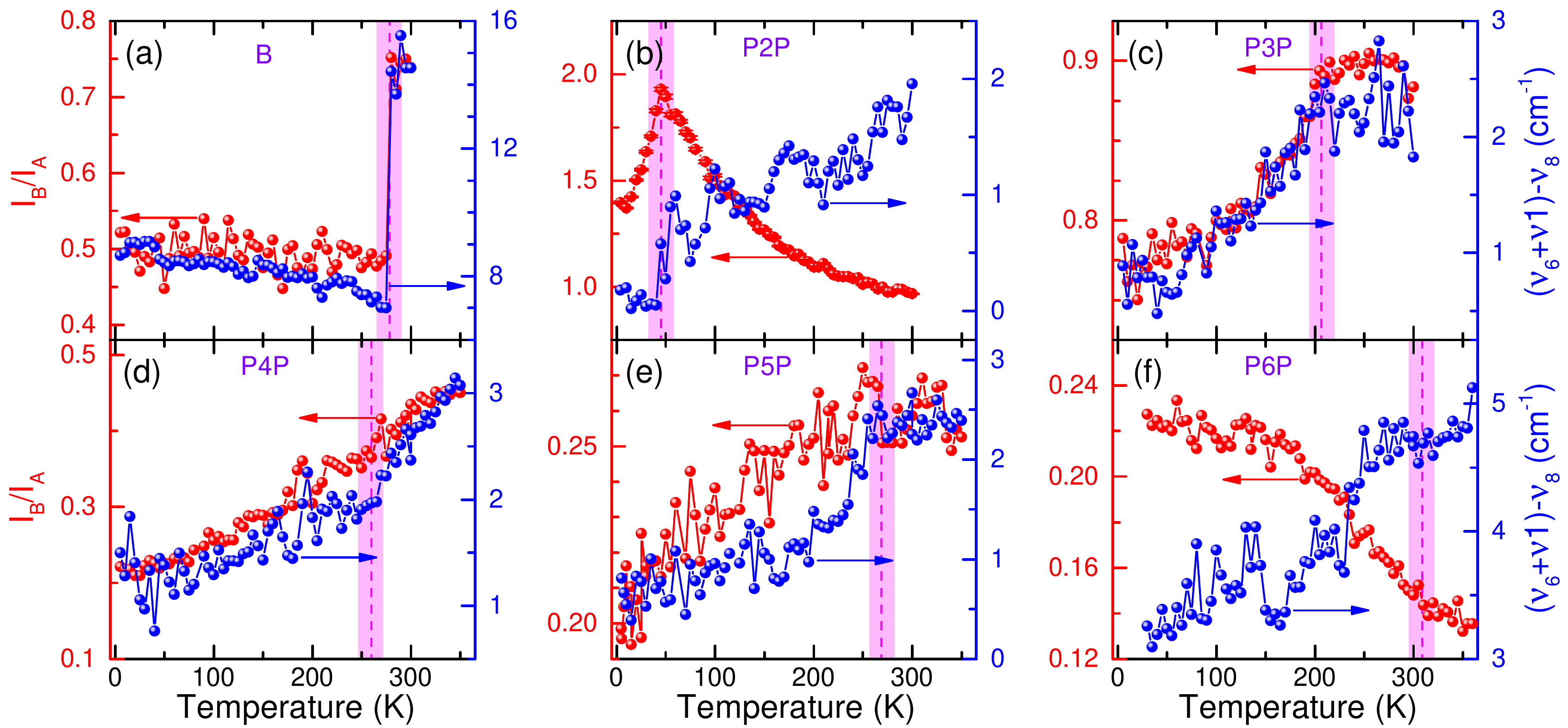}
\caption{Intensity ratios of the higher energy peak (B) and the lower energy peak (A) as a function of temperature.}
\end{figure*}

The C-H in-plane bending modes and the interring C-C stretching modes located at around 1220 and 1280 cm$^{-1}$ are common indicators to identify the molecular conjugations at ambient condition (chain length and planarity).\cite{Ohtsu,Heime,Cuff,Heime-1,Zhang-1,Marti-0} The energy separation of these two modes increases with increasing phenyl ring number in one molecule. And the intensity ratio of the POPs monotonously declines, but the pace of the declines slows, it almost converges to zero for the infinite long polymer (see Fig. 5(a)). Figure 4 presents the temperature evolutionary Raman spectra of these two modes. The frequencies of all these two peaks are almost stable at normal state, and all they exhibit broad widths, whereas they all show fierce splits below the transition temperatures which point out by dashed lines, the splits primarily result from the drastic decreases of the FWHMs of each mode which can be clearly seen in the maps. Interestingly, the 1220 cm$^{-1}$ modes of the four POPs all red-shift with decreasing temperature, whereas the 1280 cm$^{-1}$ modes blue-shift, except the P2P which the 1280 cm$^{-1}$ mode red-shifts upon cooling. According to the earlier theoretical work, the former mode (see Fig. 4(a)) is rationalized to be associated to the delocalized band which stems largely from the C atoms on the \emph{para} positions, and the latter mode is more determined by the localized band results from the off-axis C atoms.\cite{Heime-1} Consequently, with decreasing temperature, the energy of the delocalized band decreases, and the localized band has the opposite change. In addition, below the transition temperature, the twist angles appear between the neighboring phenyl rings. Thus, both the delocalized and localized bands will have anomalous changes.

The intensity ratios of the POPs are presented in Fig. 4 evaluated by the right axes. The intensity ratio is always used to indicate the molecular conjugation and the structural transition.\cite{Ohtsu,Heime,Cuff,Heime-1,Marti-0,Zhang-1,Zhang-2,Zhang,Zhang-3} However, all these two peaks exhibit severe splits below the transition temperature, it makes the selects of the two peaks more difficult. Here, we choose two peaks with strongest intensities at around 1220 and 1280 cm$^{-1}$, respectively, and the results are compared with the P2P, P3P, and P4P. All the ratios show anomalous changes at around the structural transition temperatures, but the amplitudes of the anomalies gradually decline with increasing chain length, and in the P6P, any anomalous signs are hardly seen to identify the transition temperature. It shares the same tendency with the FWHM of the lowest energy peak. Such a phenomenon should also result from the decrease of the inter-ring torsion angles of the longer polymers.

\begin{figure}[htbp]
\includegraphics[width=0.48\textwidth]{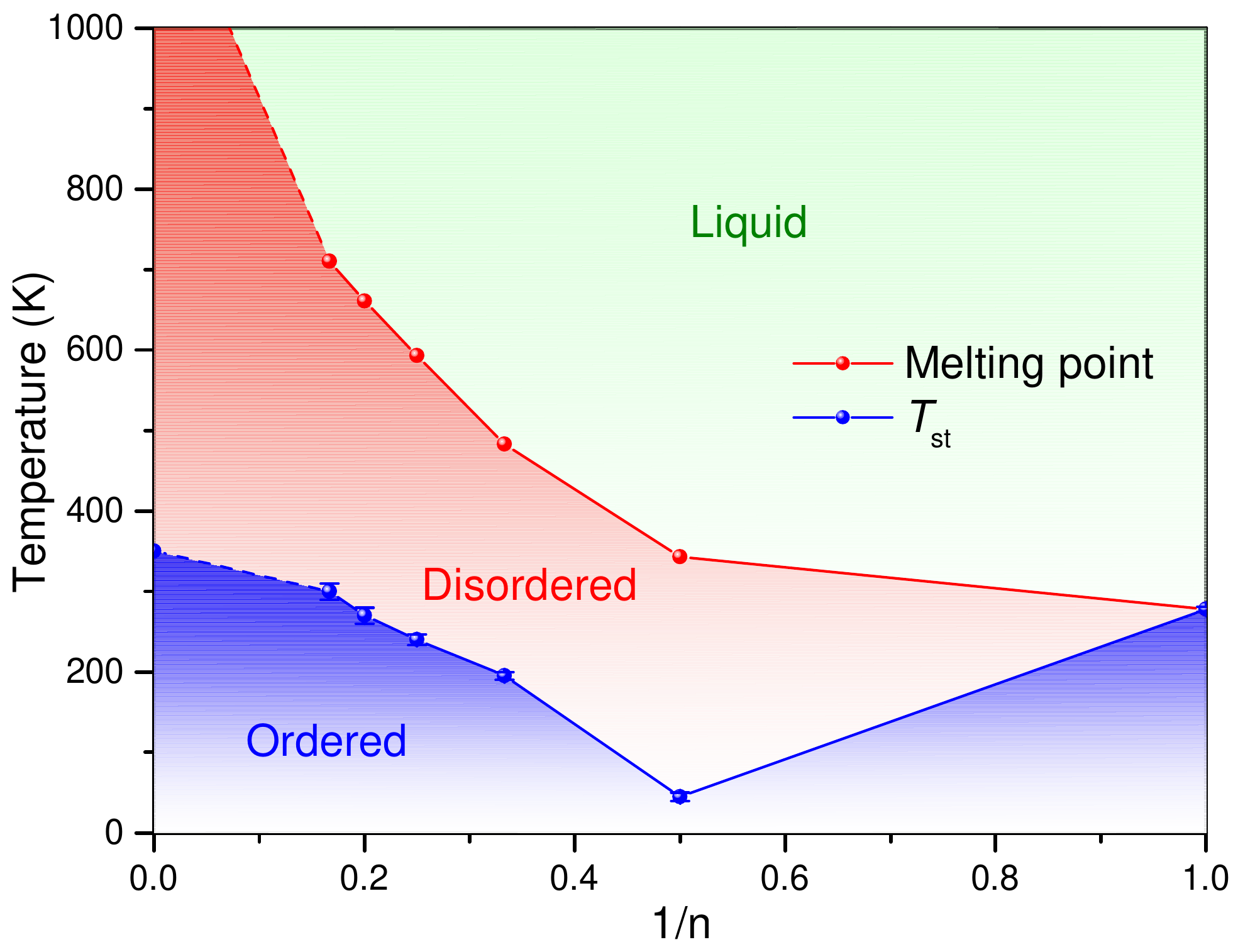}
\caption{Phase diagram of the melting point and the structural transition temperature of benzene, biphenyl, $p$-terphenyl, $p$-quaterphenyl, $p$-quinquephenyl, and $p$-sexiphenyl. $T$$_{\rm st}$: The structural transition temperature.}
\end{figure}

The temperature evolutionary intensity maps of modes at around 1600 cm$^{-1}$ of benzene and POPs are presented in Fig. 5. Two strong peaks appear at room temperature, whereas they split in several peaks at low temperatures. These splits result from the reductions of the intensities and the FWHMs with decreasing temperature, and thus the weak peaks appear at low temperature. In the maps of benzene, all the two peaks exhibit broad widths and weak intensities. After crystallizing, there are abrupt drops in the widths of these two peaks, as well as the frequencies. The intensities also have sudden increases below the crystallization temperature. In addition, those anomalies occur almost at all the vibrational modes. All the changes should result from the benzene crystallizes into an orthorhombic structure from a disorder state. The anharmonic effect is significantly reduced due to the ordered arrangement of the molecules after crystallization. Meanwhile, because of the greater molecular force field surrounding the benzene molecules, the benzene rings will be swelled. This will effectively reduce the vibrational energies among the atoms, $i.e.$, the peaks in the spectra will have redshifts. Such a phenomenon makes the Raman spectroscopy a powerful tool to estimate the melting point of benzene, this method is also applicable to other materials. For the spectra of the POPs, all those two peaks blue-shift upon cooling, and show few anomalies at around the transition temperatures except the P2P, the frequencies, widths and intensities of these two modes of P2P all exhibit unusual tendencies below 45 K. This should result from the typical "displacive" type transition which is distinct from the "order-disorder" type transition of other POPs. The intensity ratio of these two modes are also always used to estimate the molecular conjugation.\cite{Ohtsu,Zhang-1,Heime-3} However, due to the drastic splits of these peaks, it becomes more difficult to distinguish the peaks. Here we choose two peaks located at around 1600 cm$^{-1}$ with the strongest intensities, then calculate the intensity ratios of the five materials, the temperature dependent intensity ratios between the lower energy peaks and the higher energy peaks of POPs are presented in Figs. 6(e) and (f), and compared with the P2P, P3P, and P4P.\cite{Zhang-2,Zhang,Zhang-3} All the ratios show anomalous variations below the transition temperatures. However, the anomalies become increasingly obscure with increasing chain length. One can see clear anomalies in P2P and P3P, whereas the other materials show few anomalies, this should also result from the gradually decreased torsion angles of the longer polymers.

The theoretical calculation of these two peaks suggested that they stem from the "resonance split" of one peak.\cite{Heime-3,Wilso,Angus,Ito} The lower energy peak A is a fundamental band, and the higher energy peak B ($\nu$$_8$) is a combination tone of the two fundamental bands $\nu$$_1$ and $\nu$$_6$ respectively at around 992 and 606 cm$^{-1}$. In addition, the energy separation between the $\nu$$_8$ and $\nu$$_6$+$\nu$$_1$ is an indicator of the mode mixing level. It also has responses to both the inter-ring twist and chain length, and thus can be used to estimate the molecular conjugation.\cite{Heime-3,Zhang-2,Zhang,Zhang-3} The calculated ($\nu$$_6$+$\nu$$_1$)-$\nu$$_8$ of benzene and POPs are shown in Fig. 6. The energy separation of benzene shows different feature, it suddenly drops below the crystallization temperature, and then increases with decreasing temperature. This should result from the combination of the chain length effects and the inter-ring twists. All the energy separations of POPs show descending tendencies upon cooling. It indicates that the mixing levels between the fundamental and combination bands decline at low temperatures. In addition, they also have anomalous changes at around the transition temperatures. It implies that the mixing is also sensitive to the ring twist.

In the end, a corresponding phase diagram of the melting point and structural transition temperature is plotted in Fig. 8. It is used as a guide to predict the physicochemical properties of the polymer with long chain length. In this figure, the melting point monotonously increases from benzene to \emph{p}-sexiphenyl, and the extrapolated temperature of the infinite long polymer is more than 1000 K. This is identical to the previous works which suggests that the infinite long PPP is almost infusible.\cite{Baker} The structural transition temperature also increases with increasing chain length, whereas the growth rate gradually decreases. Benzene is suggested to have no any structural transition at low temperatures, \emph{i.e.}, in the crystalline benzene, all the benzene rings should be parallel to each other. The extrapolated temperature of the infinite long polymer is about 350 K.

\section{Conclusions}

In conclusion, Raman scattering spectra of benzene, \emph{p}-quinquephenyl, and \emph{p}-sexiphenyl at different temperatures were performed. The vibrational properties of those materials were comprehensively analyzed, and the results were compared with the previous works focused on the biphenyl, \emph{p}-terphenyl, and \emph{p}-quaterphenyl. All these three indicators of the molecular conjugation, including the lowest energy peak and the intensity ratios between the 1280 cm$^{-1}$ and 1220 cm$^{-1}$ modes and between the two modes at around 1600 cm$^{-1}$, were examined at low temperature condition. The frequency of the lowest energy peak remains the declining tendency with increasing chain length. Meanwhile, all the \emph{p}-oligophenyls exhibit drastic drops on the peak widths of the lowest energy peaks, and thus can be used to estimate the molecular conjugation at low temperatures. The other two indicators also have anomalous changes at around the structural transition temperatures, and are also applicable in the low temperature condition. However, the drastic splits of those modes make the calculations of the intensity ratios more difficult. In addition, due to the gradual decrease of the inter-ring twist with increasing molecular chain length, the amplitudes of the anomalies gradually vanish. It is almost indistinguishable in the \emph{p}-sexiphenyl and the longer polymers. In the end, a phase diagram of the melting point and the structural transition temperature was estabilished. They all increase with increasing chain length, and the extrapolated melting point and transition temperature for the polymer with infinite chain length are above 1000 K and 350 K, respectively.


This work was supported by the National Key R$\&$D Program of China (Grant No. 2018YFA0305900).

\end{document}